\documentclass[aps,prl,twocolumn,showpacs,superscriptaddress,groupedaddress]{revtex4}  
\usepackage{graphicx}  
\usepackage{dcolumn}   
\usepackage{bm}        
\usepackage{amssymb} 
\usepackage{amsmath}
 
\newcommand{\be}{\begin{equation}}
\newcommand{\ee}{\end{equation}}
\newcommand{\bea}{\begin{eqnarray}}
\newcommand{\eea}{\end{eqnarray}}

\begin{document}

\title{Self-similar equilibration of  strongly interacting systems from holography}
\affiliation{Department of Physics, School of Applied Mathematics and Physical Sciences, National Technical University, 15780 Athens, Greece}
\affiliation{STAG Research Centre and Mathematical Sciences, University of Southampton, Southampton, UK}
\affiliation{School of Mathematical Sciences, Queen Mary University of London, Mile End Road, London E1 4NS, UK}
\affiliation{DAMTP, Centre for Mathematical Sciences, Wilberforce Road, Cambridge CB3 0WA, UK}

\author{Ioannis Bakas} \affiliation{Department of Physics, School of Applied Mathematics and Physical Sciences, National Technical University, 15780 Athens, Greece}
\author{Kostas Skenderis} \affiliation{STAG Research Centre and Mathematical Sciences, University of Southampton, Southampton, UK}
\author{Benjamin Withers} \affiliation{STAG Research Centre and Mathematical Sciences, University of Southampton, Southampton, UK} \affiliation{School of Mathematical Sciences, Queen Mary University of London, Mile End Road, London E1 4NS, UK} \affiliation{DAMTP, Centre for Mathematical Sciences, Wilberforce Road, Cambridge CB3 0WA, UK}

\date{\today}

\begin{abstract}
We study the equilibration of  a class of far-from-equilibrium strongly interacting systems using gauge/gravity duality.
The systems we analyse are 2+1 dimensional and have a four dimensional gravitational dual.
A prototype example of a system we analyse is the equilibration of a two dimensional fluid which is translational invariant in one direction and is attached to two different heat baths with different temperatures at infinity in the other direction. We realise such setup in gauge/gravity duality by 
joining two semi-infinite asymptotically Anti-de Sitter (AdS) black branes of different temperatures, which subsequently evolve towards equilibrium by emitting gravitational radiation towards the boundary of AdS. 
At sufficiently late times the solution converges to a similarity solution, which is only sensitive to the left and right equilibrium states and not to the details of the initial conditions. 
This attractor solution not only incorporates the growing region of equilibrated plasma but also the outwardly-propagating transition regions, and can be constructed by solving a single ordinary differential equation.
\end{abstract}

\maketitle


Far-from-equilibrium dynamics is a topic of considerable interest, yet it is theoretically poorly understood, particularly for strongly interacting systems which do not admit a quasi-particle picture. In this Letter we aim to use gauge/gravity duality to study the equilibration process for a class of strongly interacting systems. 
Through the duality, thermal states are dual to stationary AdS black holes. Away from equilibrium, the system is described by the evolution of the Einstein equations subject to appropriate boundary conditions at the AdS boundary.

There has been a strong interest in the application of holographic techniques to out of equilibrium phenomena, including examples in thermalisation~\cite{Balasubramanian:2011ur}, heavy ion collisions \cite{Janik:2010we,Chesler:2015wra}, turbulence \cite{Adams:2013vsa}, dynamical and stationary quenches in normal and superfluid phases \cite{Bhaseen:2012gg,Figueras:2012rb}, to name a few. Typically numerical techniques are required to evolve the Einstein equations, but with certain simplifying assumptions simpler models can be used to shed some light on the underlying physics, for example, the use of the analytic Vaidya spacetime in the context of thermalisation \cite{Bhattacharyya:2009uu}.

In this Letter we study the evolution of 2+1 dimensional systems (thin films) which are translational invariant in one direction and are attached to conformally invariant heat baths with different temperatures in the other direction. There has been a recent interest in the study of such configurations, with a universal steady-state flow conjectured to emerge in the equilibrating region \cite{Bernard:2012je, 2013JPhA...46K2001B, Chang:2013gba, Bhaseen:2013ypa, Bernard:2015bba}. Explicit constructions of gravitational dual solutions show agreement \cite{Amado:2015uza} with the proposed ansatz.

We shall show that this setup can be captured by the Robinson-Trautman (RT) class of solutions to the Einstein Equations \cite{RT}. The RT spacetimes are 4d solutions which can be constructed by solving a single 3d parabolic partial differential equation (PDE) for a field $\sigma$, which includes the asymptotically AdS spacetimes of interest here. Remarkably, the equation that one gets is qualitatively similar to that appeared in earlier studies of thin films (see \cite{RevModPhys.81.1131} for a review).

By choosing appropriate boundary conditions for $\sigma$, we can set up the situation described above, wherein we join two different thermal states and study the evolution towards equilibrium. Remarkably, we find that at sufficiently late times the process of equilibration is only sensitive to the left and right thermal states, and becomes independent of any other details of the initial conditions. The universal solution is similarity-invariant, and is governed by an ordinary differential equation (ODE).  The use of RT solutions in the context of AdS/CFT has been studied in \cite{deFreitas:2014lia, Bakas:2014kfa}.

One important aspect of the RT solutions is that the boundary metric is inhomogeneous and time-dependent, in concert with the state of the quantum system. This can be viewed as resulting from deformations of the CFT which ensure that the bulk metric is of RT type. In other words, the Hamiltonian of the system is time-dependent and becomes conformally invariant at late times.
As such, our results are not expected to agree directly with those of \cite{Bernard:2012je, 2013JPhA...46K2001B, Chang:2013gba, Bhaseen:2013ypa, Bernard:2015bba,Amado:2015uza}. Nevertheless, we have described a new class of universal late-time gravitational behaviour governing far from equilibrium CFTs with deformations; it would be interesting to see whether such self-similar phenomena has wider applicability, for instance beyond the scope of the RT solutions themselves and to other CFT settings where the deformations may be  better understood.


{\em Robinson-Trautman.}--- The RT solutions can be viewed as a nonlinear generalisation of algebraically special perturbations of 4d Schwarzschild black holes which describe purely outgoing gravitational radiation. These perturbations can be lifted to a nonlinear solution, resulting in a time-dependent, inhomogeneous spacetime possessing a shear-free, irrotational null geodesic congruence. 
RT solutions exist for any value of the cosmological constant, $\Lambda$. Here we take the case of asymptotically locally AdS spacetimes, with $\Lambda = -3/L^2$. In retarded-time, $u$, the line element is given by
\bea
ds^2 &=& -F du^2 - 2 du dr + \frac{r^2}{\sigma^2}(dx^2+dy^2) \label{RTplanar}\\
F &\equiv& -\frac{\Lambda}{3}r^2  -2 r\frac{\partial_u \sigma}{\sigma}+ \sigma^2\nabla_{R^2}^2 \log\sigma- \frac{2m}{r},
\eea
where $\nabla^2_{R^2}$ is the standard Laplacian for $R^2$. This line element satisfies Einstein equations
provided $\sigma(u,x,y)$ satisfies a certain fourth-order nonlinear parabolic equation on R$^2$. This equation can be phrased as a geometric flow; let us define an Euclidean 2-metric
\be
 \gamma_{ij}(u,x,y)dx^i dx^j = \frac{1}{\sigma^2(u,x,y)} (dx^2+dy^2),
\ee
then $\gamma$ obeys the Calabi flow equation, i.e.
\be
\partial_u \gamma_{ij} = \frac{1}{12m} \nabla^2_\gamma R_\gamma\, \gamma_{ij} \label{calabiflow}
\ee
which is now an equation for $\sigma$,  where $\nabla^2_\gamma$ and $R_\gamma$ are the 
 Laplacian and  Ricci scalar for $\gamma$. Note that this equation is insensitive to $\Lambda$. Also note that because of the application we have in mind, we have restricted to Calabi flow on R$^2$.

With $\Lambda<0$ the bulk evolution corresponding to a solution of (\ref{calabiflow}) is holographically dual to an out of equilibrium CFT on a time dependent, inhomogeneous background metric $g$, given by,
\be
g_{\mu\nu} dx^\mu dx^\nu =  -dt^2 + \gamma_{ij}(t,x,y)dx^i dx^j, \label{boundarymetric}
\ee
where we have introduced the time coordinate $t$ which is given by the value of $u$ at the conformal boundary.
For constant $\sigma$ the bulk geometry is the Schwarzschild black brane solution, and the CFT is in thermal equilibrium on Minkowski space. 

In this work we have taken the spatial part of the metric (\ref{boundarymetric}) to be non-compact. For compact cases, given smooth initial data for $\sigma$, the solution converges to a constant at late times with corrections which vanish exponentially fast with $u$ \cite{Schmidt, Rendall, Singleton, Chrusciel:1992cj, Chrusciel:1992rv}. Thus in the compact case, the system settles down to the equilibrium Schwarzschild solution. This result does not apply to our planar solutions. In fact, we will see that depending on boundary conditions at spatial infinity, the system converges to a time-evolving similarity solution with polynomial corrections at late times.


{\em Similarity.}--- The Calabi flow equation (\ref{calabiflow}) is a fourth-order parabolic PDE, which as we shall show, admits similarity solutions that play an important role in the nonlinear dynamics of the CFT.
First, as a warm-up example of similarity in such systems, consider instead the heat equation,  $\tfrac{\partial f}{\partial t} = D \tfrac{\partial^2 f}{\partial x^2}$, where $D$ is the thermal diffusivity. This admits solutions with the scaling symmetry, $x\to \lambda^2 x, t\to \lambda t$, which are easily obtained by writing $f$ as the function of a single invariant variable, $f(t,x) = h(\mu(x,t)) \quad \text{where} \quad \mu(t,x) = x/\sqrt{t}$. In doing so the equation is reduced to an ODE, and one solution is
\be
h = a+b\, \text{erf}\left(\frac{\mu}{2\sqrt{D}}\right)
\ee
 where erf is the error function and $a,b$ are integration constants. $h$ becomes a constant as $x\to\pm\infty$ for fixed $t$. This solution corresponds to the evolution resulting from initially joining two semi-infinite systems of different temperatures; at $t=0$ the solution is a step function centred on $x=0$ where the two infinite systems are initially joined.

We can analogously look for similarity solutions of the Calabi flow equation, which if they exist for similar boundary conditions, describe the evolution resulting from a particular way of connecting two equilibrium black brane solutions. The two different black branes should have different temperatures (and hence different masses) and this corresponds to having $\sigma$ approach different values at infinity.

As a first step we note that the RT metric (\ref{RTplanar}) is invariant if we make the following scalings, 
\bea
u &\to& \lambda_u u, \quad x \to \lambda_x x,\quad y\to\lambda_x y \nonumber\\
r &\to& \lambda_u^{-1} r,  \quad m \to \lambda_u^{-3} m, \quad \sigma \to \lambda_x\lambda_u^{-1} \sigma.\label{RTScalings}
\eea
For $m=0$ this corresponds generically to a Lifshitz scaling isometry.
As expected however, at finite temperature the symmetry is broken by $m$, which itself must scale appropriately. 
We do not consider the $m=0$ limit here and so we will always be in the broken setting; the algebraically special modes are non-analytic at $m=0$, and we return to this limit in \cite{longpaper}.

We now impose translational invariance in the $y$ direction (so in particular $\sigma$ is independent of $y$) and
as with the heat equation example we seek solutions which are manifestly scale invariant. As we shall see momentarily, an ansatz appropriate for (\ref{RTScalings}) is
\be
\sigma(t,x) =  m^{1/4} (t-t_0)^p h(\mu(t,x)),\label{simansatz}
\ee
where $\mu(t,x) \equiv (x-x_0)/(t-t_0)^{p+\frac{1}{4}}$ and $t_0$ and $x_0$ are parameters corresponding to time translations and spatial translations respectively, and correspond to the location of the `join'. 
We have allowed for an additional parameter $p$, extending the family of similarity solutions.
For this ansatz (\ref{calabiflow}) becomes an $m$-independent ODE,
\be
\partial_\mu^4h = \frac{(\partial_\mu^2 h)^2}{h} + 3(1+4p) \mu  \frac{\partial_\mu h}{h^4} - \frac{12p}{h^3}. \label{heqn}
\ee
Note that this equation has a scaling symmetry,
\be
h\to \lambda h, \quad \mu \to \lambda \mu. \label{hscaling}
\ee
Compatibility with the bulk scaling property (\ref{RTScalings}) requires $\lambda=\lambda_x/\lambda_u^{p+1/4}$. 
As we shall see below, explicit solutions $h$ of (\ref{heqn}) do not transform as $\mu \to \lambda \mu$ and thus $\lambda$ must be equal to one (so that (\ref{hscaling}) holds identically).  This then fixes the Lifshitz dynamical critical exponent to be $z = \left(p+1/4\right)^{-1}$. 
For concreteness we will now focus on the case $z=4$, $p=0$ -- we find similar behaviour for $p\neq 0$ solutions, which we turn to at the end.

\begin{figure}[h!]
\begin{center}
\includegraphics[width=0.48\textwidth]{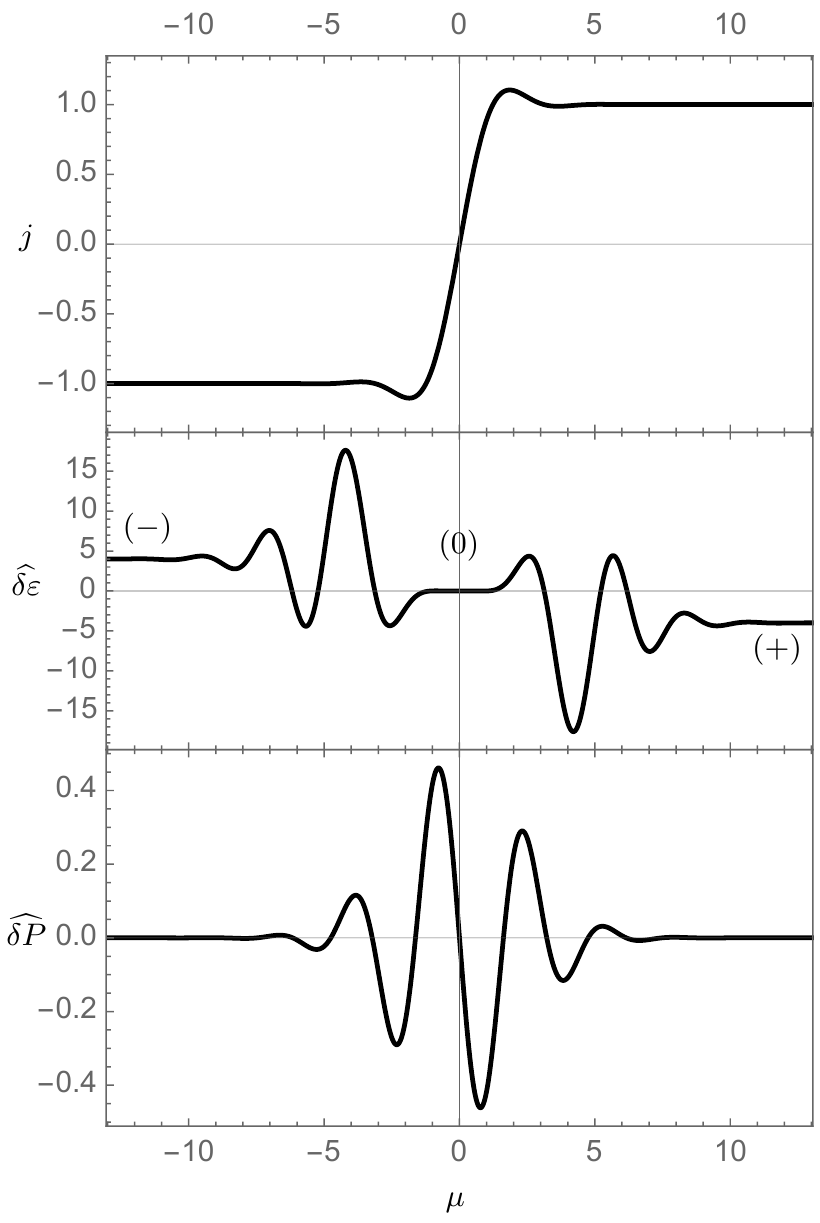}
\caption{The similarity solution describing closely separated equilibria, after performing a certain Weyl transformation. \emph{Top:} The function $j$ describing the deviation from equilibrium for the metric function $\sigma$. \emph{Middle:} The energy density obtained from the eigenvalue equation, $\widehat{\delta \varepsilon}$ which shows the deviation from equilibrium and is defined in the text. The energy density is independent of time at fixed $x/t^{1/4}$ for each of the three flat space regions labelled by $(-),(0),(+)$. The central equilibrium region has a spatial extent which grows as $t^{1/4}$. \emph{Bottom:} The time-rescaled difference between longitudinal and transverse pressures, $\widehat{\delta P}$, also defined in the text. 
 \label{linfig}}
\end{center}
\end{figure}

To solve (\ref{heqn}) we begin by looking for solutions describing the equilibration of nearby thermal states, i.e. we linearise about the particular $p=0$ solution $h=1$, $h(\mu) = 1+ \epsilon j(\mu)$, where we have introduced a small parameter $\epsilon$. $j$ satisfies 
\be
\partial_\mu^4j = 3\mu \partial_\mu j \label{jeqn}
\ee
which admits a solution in terms of hypergeometric functions. The solution which is regular for all $\mu$ and asymptotes to a constant is given by
\be
j= \frac{-3^\frac{1}{4}4\mu}{\Gamma\left(-\frac{1}{4}\right)}\,
_1 F_3\left(^\frac{1}{4}_{\frac{1}{2},\frac{3}{4},\frac{5}{4}};\frac{3 \mu^4}{64}\right)-\frac{\Gamma\left(\frac{3}{4}\right)\mu^3}{\sqrt{2} 3^\frac{1}{4} \pi}\; _1 F_3\left(^\frac{3}{4}_{\frac{5}{4},\frac{3}{2},\frac{7}{4}};\frac{3 \mu^4}{64}\right) \label{analyticj}
\ee
Here  $\lim_{\mu\to-\infty} j = -1$, $\lim_{\mu\to\infty} j = 1$ and $j(0)=0$. This is the general solution at this order in perturbations; other constant boundary conditions can be reached using linearity and shift symmetry. We plot $j$ in the top panel of FIG. \ref{linfig}. 

\begin{figure}[h]
\begin{center}
\includegraphics[width=0.48\textwidth]{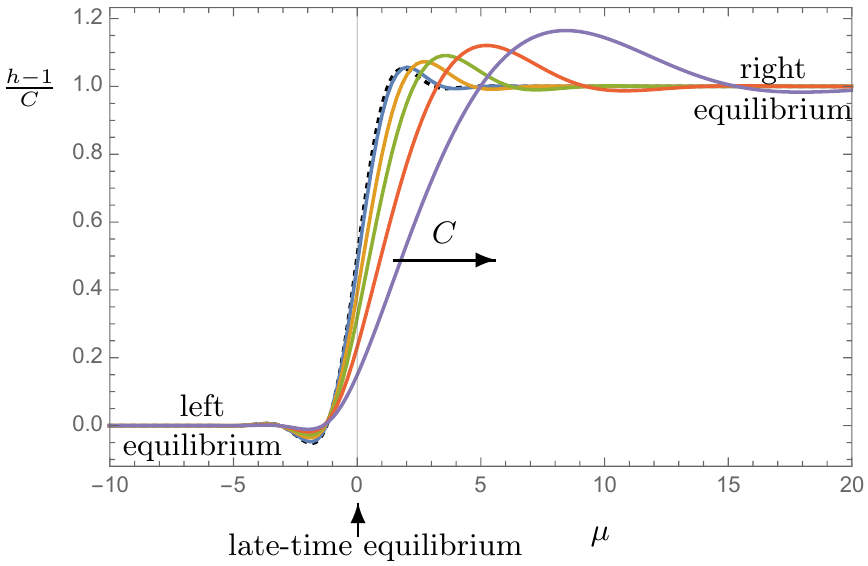}
\caption{Cohomogeneity-1 similarity solutions to the Calabi-flow equation on the plane. The coordinate $\mu$, defined in (\ref{simansatz}), is a scale-invariant quantity, and so the spatial profile is expanding with time. The solid curves show different values of $C$ which label the equilibrium state of the right hand asymptotic system as a Dirichlet boundary condition (\ref{dirichletbcs}). The values from left to right are $C = 0.1,0.5,1.0,2.0,4.0$ and the dashed curve is the linear solution (\ref{analyticj}). \label{simfig}}
\end{center}
\end{figure}

To go to widely separated left and right thermal states, we proceed numerically.  In detail, we use 6th order finite differences in a compactified spatial coordinate, $R = \tanh\left(\frac{\mu}{\ell}\right)$ where $\ell$ is chosen so that a uniform grid in $R$ usefully covers the region in $\mu$ where $h$ is varying significantly. For the examples below we have taken $\ell=20$. The system is solved using a Newton-Raphson method, giving Dirichlet boundary conditions at $R = \pm 1$ corresponding to the constant asymptotic values of $h$. For concreteness we fix,
\be
h(R = -1) = 1, \quad h(R = 1) = 1+C. \label{dirichletbcs}
\ee
Any other pair can be brought into this form using the symmetry (\ref{hscaling}). Some solutions are shown in FIG. \ref{simfig}, with a clear deviation from the linearised solution for sufficiently large $C$.

{\em QFT Interpretation.}--- In order to interpret the similarity solutions in the face of the inhomogeneous evolving boundary metric (\ref{boundarymetric}), on the boundary we can simultaneously perform a Weyl transformation and coordinate transformation. It is possible to do so such that for $|x|\gg t^{1/4}$ and $|x|\ll t^{1/4}$, the metric is simply $ds^2 = -dt^2+L^2(dx^2+dy^2)$. In this frame we can simultaneously discuss the equilibrium state of the system on the left and the right and observe a growing flat space region in the interior as part of an out-of-equilibrium evolution. This can be achieved through the Weyl transformation $g_{new} = \Omega^{8/3} g$ together with the coordinate transformations
\be
x^i\to x^{i'} = \frac{x^i}{\Omega^{1/3}},\qquad t\to t' = \frac{t}{\Omega^{4/3}}.
\ee
for 
\be
\Omega(\mu) = \sigma - \frac{\mu}{1!} \sigma'+ \frac{\mu^2}{2!} \sigma''- \frac{\mu^3}{3!} \sigma'''+\frac{\mu^4}{4!} \sigma'''',\label{frame}
\ee
where prime denotes derivative w.r.t. $\mu$. 

One may extract the holographic stress tensor from the asymptotics of the solution near the conformal infinity \cite{deHaro:2000vlm}. For RT solutions this has been done in \cite{deFreitas:2014lia, Bakas:2014kfa}.  The stress energy tensor may be expressed as a sum of a stress energy tensor due to a conformal perfect fluid  and a part that depends on the Cotton tensor of the boundary metric \cite{deFreitas:2014lia} \footnote{The Cotton tensor of a three-dimensional metric $\gamma_{ab}$ is defined as follows,
$
C^{ab}  =  \frac{\epsilon^{acd}}{\sqrt{- {\rm det} \gamma}} \nabla_c
\left({R^b}_d - \frac{1}{4} {\delta^b}_d R \right) ,
$
where $\epsilon^{acd} $ is the anti-symmetric symbol and $R_{ab}, R$ are the Ricci tensor and curvature scalar of the metric $\gamma_{ab}$.}.
In the left and right asymptotic regions the holographic stress tensor is simply,
\be
\kappa^2 \left<T^{\pm}_{\mu\nu}\right> = \frac{m}{\sigma_{\pm}^4}\,\text{diag}\left(\frac{2}{L^2},1,1\right)
\ee
where the $\pm$ labels values at $x\to \pm \infty$.
In the growing central region $|x|\ll t^{1/4}$ the stress tensor is
\be
\kappa^2 \left<T^{0}_{\mu\nu}\right> = \frac{m}{\sigma(0)^4}\,\text{diag}\left(\frac{2}{L^2},1,1\right) +\Pi_{\mu\nu}
\ee
where the additional term,
\begin{align}
\Pi_{\mu\nu} dx^\mu dx^\nu =& -\frac{L^2}{4} \frac{1}{t^{3/2}}\frac{\sigma''(0)}{\sigma(0)} (dx^2-dy^2) \\
&+ \frac{1}{t^{3/4}} \left(\frac{\sigma'''(0)}{\sigma(0)} - \frac{\sigma'(0) \sigma''(0)}{\sigma(0)^2}\right)dt\,dx \nonumber
\end{align}
represents he energy flow across this region and it originates from the Cotton tensor.
This contribution dilutes with time, resulting in a third equilibrium region for $|x|\ll t^{1/4}$ at sufficiently large $t$. Corrections to these expressions appear with size $O(x\, t^{-1/4})$.

We can illustrate these three regions by turning to the energy density of the linear solution describing nearby equilibria, (\ref{analyticj}). The energy density can be defined by solving the following eigenvalue problem,
\be
T^\mu_{~\nu} u^\nu = -\varepsilon u^\mu,
\ee
where $u^\mu$ is timelike unit-normed vector. For the frame defined in (\ref{frame}), 
\be
\varepsilon  = \frac{2m}{L^2\kappa^2} \left(1{-} 4 \epsilon \left(j{-}\mu j'{+}\frac{1}{2!}\mu^2 j''{-}\frac{1}{3!}\mu^3 j'''{+}\frac{1}{4!}\mu^4 j''''\right)\right).\label{linearenergy}
\ee
With $u^\mu$ known, we can define two spacelike, unit-normed orthogonal vectors $n_1$ and $n_2$ which are also orthogonal to $u$. These are chosen such that in equilibrium $n_{1}$ is the unit vector in the $x$ direction. Using these we can define pressures,
\be
P_{1} = T_{\mu\nu}n_1^\mu n_1^\nu,\quad P_{2} = T_{\mu\nu}n_2^\mu n_2^\nu.
\ee
For the linear solutions in the frame (\ref{frame}), 
\be
P_{1,2} =  \frac{\varepsilon}{2} \mp \epsilon\frac{\mu j'''+2 j''}{8 \kappa^2 t^{3/2}} \label{pressureLinear}
\ee
where $\varepsilon$ is given in \eqref{linearenergy}. The off-diagonal term $T_{\mu\nu}n_1^\mu n_2^\nu= O(\epsilon)^2$. The energy density and pressure is shown in the lower two panels of FIG. \ref{linfig}, where we plot $\widehat{\delta \varepsilon} \equiv \partial_\epsilon \varepsilon |_{\epsilon= 0}/(\varepsilon|_{\epsilon= 0})$, and $\widehat{\delta P} \equiv \epsilon^{-1}\kappa^2 t^{3/2}(P_2-P_1)$ as a function of $\mu$. We can see the emergence of the equilibrium state situated between the two reservoirs, whose spatial extent grows like $t^{1/4}$. This is the thermal state reached at sufficiently late times for any fixed $x$.

{\em Stability.}--- The similarity solutions that have been discussed hitherto, are not only simple examples of possible evolutions, but play a crucial role in determining the late time behaviour of the system. For the evolution of initial data which falls outside of the ansatz (\ref{simansatz}) the system is not initially described by a similarity solution, but does settle down to the similarity solution at sufficiently late times, as prescribed by the left and right asymptotic equilibria. As a first step to establishing this result, we consider a perturbation,  $\chi$,  to the similarity solution within the RT class which preserves the left and right asymptotics,
\be \label{powerlaw}
\sigma(t,x) = m^{\frac{1}{4}} \left(h\left(\frac{x}{t^{1/4}}\right) + \chi\left(\frac{x}{t^{1/4}},t\right)\right).
\ee
The resulting linear PDE for $\chi$ admits separable solutions with power law decay,
\be
\chi  = t^{-\Delta} \chi_\Delta\left(\frac{x}{t^{1/4}}\right)
\ee
where $\chi_\Delta$ satisfies the following eigenvalue problem, 
\bea
{\cal O} \chi_\Delta &=& \Delta \chi_\Delta \label{chieval}\\
{\cal O} &\equiv& \frac{1}{12}h^4 \partial_\mu^4 - \frac{1}{6} h^3 h'' \partial_\mu^2 - \frac{1}{4}\mu \partial_\mu + \frac{1}{12} h^2 (h'')^2 + \mu \frac{h'}{h} \nonumber
\eea
The scaling (\ref{powerlaw}) suggests that the dynamics of the system is governed by a Lifshitz invariant critical point with dynamical exponent $z=4$, with the set of $\chi$s associated with spectrum of operators of this (non-relativistic) scale invariant theory.

The operator ${\cal O}$ admits a zero mode, $\chi_{0} = h-\mu h'$ which follows from the invariance noted earlier (\ref{hscaling}). This however does not preserve the boundary conditions and so can be excluded from the late time spectrum. The spectrum about an $h=1$ background solution can be built in reference to the perturbative solutions $j$, (\ref{analyticj}). The equation that $\chi_\Delta$ satisfies does not depend on $j$ at order $\epsilon^0$,
\be
\frac{1}{12} \partial_\mu^4 \chi_\Delta - \frac{1}{4}\mu \partial_\mu \chi_\Delta = \Delta \chi_\Delta + O(\epsilon) \label{chih1}
\ee
Nevertheless, a set of solutions to (\ref{chih1}) which respect the boundary conditions are generated by solutions $j$, i.e.
\be
\chi_{n/4} = \partial_\mu^n j + O(\epsilon) \qquad n \in \mathbb{Z}^+.
\ee
The $n=0$ case has been excluded because it changes the asymptotics of $\sigma$ (as noted above). At least around the $h=1$ background we therefore have a spectrum of modes which is decaying, since $\Delta>0$. We may reasonably expect a positive spectrum to persist in a neighbourhood of the $h=1$ backgrounds. We have verified this by numerically computing the eigenvalue spectrum of ${\cal O}$ for the non-linear case, identifying the four longest lived modes as $\Delta = 1/4, 1/2, 3/4, 1$, invariant over a wide range of non-linear similarity solutions, $h$. Actually, for non-linear backgrounds we can construct two of these eigenfunctions exactly, $\chi_{1/4} = h'$, and $\chi_{1} = \mu h'$. These correspond to modes which translate the solution in $x$ and $t$ respectively.


{\em Numerical evidence.}--- 
We now turn to a general numerical evolution of the Calabi flow equation (\ref{calabiflow}) in the cohomogeneity-1 case, choosing initial conditions which are incompatible with the ansatz (\ref{simansatz}). For time evolution we use Crank-Nicolson, and for each implicit stage of the integration we use Newton-Raphson for 6th order finite differences with the same discretisation of the compactified coordinate $R$ as before. 

\begin{figure}[h]
\begin{center}
\includegraphics[width=0.48\textwidth]{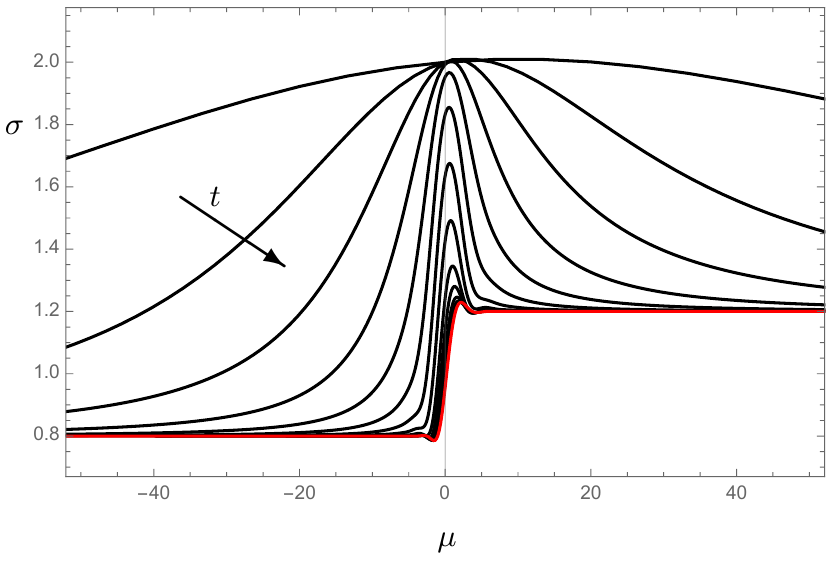}
\caption{Evolution of the initial data (\ref{egSim}) according to the equation (\ref{calabiflow}) showing convergence to the planar similarity solution in red, which is prescribed only by $\sigma_\pm$. Each curve shows a different time in the evolution.\label{universality}}
\end{center}
\end{figure}

By way of a concrete representative example in FIG. (\ref{universality}) we show the evolution of $h$ for the initial data,
\be
\sigma(0,x) = 1+\frac{2}{10} \tanh{x} + \frac{1}{1+x^2}\label{egSim}.
\ee
At late times the solutions approach the similarity solutions labelled by the left and right temperatures. 

{\em Away from $p=0$.}--- We have studied in detail the cases corresponding to $p=0$, for the primary reason that it  includes the case of a thermal state on Minkowski space. For $p\neq 0$, constant $\sigma$ is no longer a solution, and indeed $\sigma$ can become singular, but in principle these cases should not be excluded. For example, the following solution
\be
h = \sqrt{2}(1+\sqrt{2}\mu)^\frac{3}{4},\quad m=1/4\qquad p=3/4
\ee
is the planar analogue of a solution on $S^2$ which contains an AdS C-metric in the bulk, see~\cite{Stephani:624239}. It would be interesting to investigate different values of $p$ in more detail.

Additionally we note that there are similarity solutions which are rotationally invariant corresponding to a different physical setup -- we will describe this and related cases  in more detail in a forthcoming work \cite{longpaper}.

{\em Conclusions.}--- We presented a holographic study of equilibration of a class of strongly interacting systems. 
In particular, we holographically engineered 2+1 dimensional systems which at $t=0$ are described by two different thermal states infinitely separated in one direction, and then studied the subsequent evolution. It turns out that the final state is a self-similar solution which only depends on the left and right temperatures and not on the details of the initial conditions. The self-similar solutions are Lifshitz invariant and perturbations around them are governed by the spectrum of operators of an underlying Lifshitz critical theory. Our discussion should thus be applicable to all systems which are in the same universality class with this critical point.

{\em Acknowledgements.}--- 
The research of I.B. is partially supported and implemented under the ARISTEIA action of the operational programme for education and life-long learning and is co-funded by the European Social Fund (ESF) and National Resources of Greece.
K.S. is supported in part by the Science and Technology Facilities Council (Consolidated Grant ``Exploring the Limits of the Standard Model and Beyond'').
B.W.\ is supported by European Research Council grant ERC-2014-StG639022-NewNGR.

\bibliography{ooe}

\end{document}